\documentclass[a4paper,english,superscriptaddress,showpacs,twocolumn,pra]{revtex4}

\usepackage[utf8]{inputenc}
\usepackage{babel}
\usepackage{textcomp}
\usepackage{color}
\usepackage{amsfonts}
\usepackage{amsmath}
\usepackage{amssymb}
\usepackage{graphicx}
\usepackage{dsfont}

\usepackage{graphicx}

\graphicspath{{converted_graphics/}}
\begin{document}

\title{Work distribution in a photonic system}

\author{M. A. A. Talarico}
\affiliation{Departamento de F\' \i sica, Universidade Federal de Santa Catarina, CEP 88040-900, Florian\'opolis, SC,  Brazil}
\affiliation{Universidade Tecnol\'ogica Federal do Paran\'a, Campus Toledo, Rua Cristo Rei, 19 CEP 85902-490, Toledo, PR, Brazil}
\author{P. B. Monteiro}
\affiliation{Instituto Federal de Santa Catarina, Campus Florian\'opolis, 
Av. Mauro Ramos, 950, CEP 88020-300, Florian\'opolis, SC, Brazil}
\author{E. C. Mattei}
\affiliation{Departamento de F\' \i sica, Universidade Federal de Santa Catarina, CEP 88040-900, Florian\'opolis, SC,  Brazil}
\author{E. I. Duzzioni}
\affiliation{Departamento de F\' \i sica, Universidade Federal de Santa Catarina, CEP 88040-900, Florian\'opolis, SC,  Brazil}
\author{P. H. Souto Ribeiro}
\email{p.h.s.ribeiro@ufsc.br}
\affiliation{Departamento de F\' \i sica, Universidade Federal de Santa Catarina, CEP 88040-900, Florian\'opolis, SC,  Brazil}
\author{L. C. C\'{e}leri}
\email{lucas@chibebe.org}
\affiliation{Instituto de F\'{\i}sica, Universidade Federal de Goi\'{a}s, CEP 74001-970, Goi\^{a}nia, GO, Brazil}

\begin{abstract}
We present a proposal of a set-up to measure the work distribution of a process acting on a quantum system emulated by the transverse degrees of freedom of classical light. Hermite-Gaussian optical modes are used to represent the energy eigenstates of a quantum harmonic oscillator prepared in a thermal state. The Fourier transform of the work distribution, or the characteristic function, can be obtained by measuring the light intensity at the output of a properly designed interferometer. The usefulness of the approach is illustrated by calculating the work distribution for a unitary operation that displaces the linear momentum of the oscillator. Other types of processes and quantum systems can be implemented with the same scheme.
We also show that the set-up can be used to investigate the energy distribution for open dynamics described by completely positive maps. We discuss the feasibility of the experiment, which can be realized with simple linear optical components.
\end{abstract}

\maketitle

\section{Introduction}

Recent developments in the intersection of thermodynamics, information theory and quantum mechanics, generated an increasing interest in the non-equilibrium behavior of small systems, specially in what concerns the applicability and meaning of the second law of thermodynamics \cite{Jarzynski_rev,Esposito,Campisi,John}.
 
The second law states that, if a classical system is driven from an initial equilibrium state by means of a given process, the work $\mathcal{W}$ done on the system must obey the relation $\mathcal{W} \geq \Delta F$, with $\Delta F$ being the difference in the free energies between the final and the initial equilibrium states. This is valid in the thermodynamic limit, where fluctuations are suppressed, and is independent of the underlying microscopic theory. When the number of degrees of freedom decreases, quantum and classical fluctuations come into play and we expect to observe violations of this relation. However, we also expect that the second law must be obeyed on average, i.e. $\langle\mathcal{W}\rangle \geq \Delta F$, where the average is taken over many repetitions of the same process.

At this level, quantities like work and heat must be described by probability distributions and in order to obtain these distributions, two projective energy measurements must be performed on the system. One before and another after some particular process of interest takes place. To be specific, let us consider the following protocol. A system $\mathcal{S}$, whose Hamiltonian is $\mathbf{H}_{\mathcal{S}}(t)$ and initially in the thermal state $\rho^{I}_{\mathcal{S}}$, is driven by an external agent to the final state $\rho^{F}_{\mathcal{S}}$ by means of a unitary transformation $U(t)$. During the process, an amount $\mathcal{W}$ of work is done on the system. The microscopic work performed on the system in each run is defined as \cite{Kurchan,Tasaki,PRX_Jarzynski}
\begin{equation}
\mathcal{W} = \varepsilon^{F}_{m} - \varepsilon^{I}_{n},
\label{1}
\end{equation} 
where $\varepsilon^{I}_{n}$ and $\varepsilon^{F}_{m}$ are the results of energy measurements at the beginning and at the end of the process, respectively. 

Considering this protocol, the probability that one finds the system in the $m$-th eigenstate of the final Hamiltonian given that it was in the $n$-th eigenstate of the initial Hamiltonian is
\begin{eqnarray}
p_{m,n} = \frac{e^{-\beta \varepsilon^{I}_{n}}}{Z^{I}}|\langle \phi^{F}_{m}| U(t) |\phi^{I}_{n}\rangle|^{2} \nonumber \\
\equiv p_{n}|\langle \phi^{F}_{m}| U(t) |\phi^{I}_{n}\rangle|^{2},
\label{eq:prob-work}
\end{eqnarray} 
where $\beta$ is the inverse temperature (Boltzmann constant is equal to one) of the initial state, $U(t)$  is an operator representing the process and $\lbrace\vert\phi^{I(F)}_{m}\rangle\rbrace$ is the set of eigenvectors of the system initial (final) Hamiltonian, and $Z^{I}$ is the initial partition function. With this definition we can readily write the probability density of work distribution
\begin{equation}
P\left(\mathcal{W}\right) = \sum_{m,n}p_{m,n}\delta\left[\mathcal{W} - \left(\varepsilon_{m}^{F} - \varepsilon_{n}^{I}\right)\right],
\label{eq:work}
\end{equation}
from which we can compute the mean value $\left \langle \mathcal{W} \right \rangle$. 

Despite its importance, experimental investigations of this relation for classical \cite{Liphardt,Exp3,Collin,Douarche} and quantum systems  \cite{An,Batalhao} are rare. In the classical case the difficulty arises because we need to control the system whose energy is of the order of the thermal fluctuations, while in the quantum case it appears mainly due to the necessity of performing two projective energy measurements on the system. However, a new idea that offers a way to avoid these measurements came out recently \cite{Mauro,Vlatko}, allowing the investigation of such relations in a nuclear magnetic resonance setup \cite{Batalhao}. The idea relies on the reconstruction of the characteristic function
\begin{equation}
G\left(s\right) = \sum_{m,n}p_{m,n}e^{is(\varepsilon^{F}_{m} - \varepsilon^{I}_{n})}.
\label{eq:char11}
\end{equation}
which is defined as the Fourier transform of the work distribution, Eq. (\ref{eq:work}), by encoding the information about the energy eigenvalues in the phases of an interfering system in a suitably designed interferometer. Such phases can then be measured in the form of oscillations at the output.

Here, we study the work distribution for a process acting on a quantum harmonic oscillator
emulated by an optical system. The emulation of other kinds of systems and the implementation of
a variety of processes is discussed in the final part of the paper.
Light beams prepared in Hermite-Gaussian modes are analog to the energy states of the quantum harmonic oscillator \cite{Paraxial2}. The statistical mixture of these modes with proper weights is equivalent to preparing an analog system in the thermal equilibrium state. This analogy comes from the equivalence between the paraxial wave equation and the 2-D Schr\"{o}dinger equation. We show that preparing these light beams in a thermal state, and sending them through an interferometer, it is possible to measure the characteristic function corresponding to the work distribution due to a process acting on the system. The work is done through a unitary process implemented by the propagation of light through a linear optical device inside the interferometer. We present an example of process that illustrates the method. We also generalize the set-up to include the study of open system dynamics, and to measure the energy (work and heat) distribution for any completely positive map. The simplicity of this scheme allows a high level of control and the study of energy distributions that would be eventually hard to treat theoretically. It is also a candidate to extrapolate some theoretical limits, by studying for instance the transition from the quantum to the classical regime.  

\section{Experimental set-up and protocol}

\subsection{The system}

In this work we are concerned with paraxial fields. An intuitive way of defining paraxial fields is given by geometric optics, where light is represented by rays. Paraxial rays are those that lie at small angles to the optical axis of the system under consideration. In physical optics, a paraxial field can be described as $\mathcal{A}\left(x,z\right) = \Psi\left(x,z\right)e^{ikz}$, with $z$ being the direction of propagation and $k$ the wavenumber. For simplicity, we only consider one transverse direction, here denoted by $x$. The generalization for two dimensions is straightforward by replacing $x$ with a vector $\vec{r} = x \hat{i} + y \hat{j}$. Therefore, $\Psi\left(x,z\right)$ describes the field in the transverse direction at longitudinal points $z$. The Helmholtz paraxial equation describes the propagation of light in this approximation, and can
be written as \cite{Marcuse} 
\begin{equation}
\frac{i}{k}\frac{\partial \Psi\left(x,z\right)}{\partial z} = \left[-\frac{1}{2k^{2}}\frac{\partial^{2}}{\partial x^{2}} + \frac{\Delta n(x)}{n_{0}}\right]\Psi\left(x,z\right),
\label{eq:helmholtz}
\end{equation}
where $\Delta n(x)/n_{0}$ is a transverse spatial modulation of the index of refraction. We can see that Eq. (\ref{eq:helmholtz}) is analog to the time dependent Schr\"{o}dinger equation if we identify $\Psi\left(x,z\right)$ with the wave function and $\Delta n(x)/n_{0}$ as the potential $V(x)$. In this picture, the propagation along the optical axis $z$ plays the role of time evolution and the wavelength of light $\lambda = 2\pi/k$ the role of Planck's constant $h$.
Therefore, by properly modulating the index of refraction we can implement some specific potential and then study the analog quantum system. 

Here we will use this analogy to demonstrate that it is possible to implement the protocol described in Refs. \cite{Mauro,Vlatko} for the determination of the work distribution using an optical system. According to this scheme, one must have some work performed on the system and also free evolutions. The work can be implemented for the case of optical modes by means of propagation through some linear optical device (a phase mask), corresponding to a unitary transformation of the modes. However, the free evolution is not achieved with free propagation, which actually realizes evolution without the potential $V(x)$. The free evolution is implemented with an optical
transformation, so that the optical mode only acquires a global phase.
We analyse in detail the case where Hermite-Gaussian (HG) modes emulate the states of a quantum harmonic oscillator (QHO), and discuss briefly the extension to Laguerre-Gaussian modes. An interesting way of understanding the analogy between HG modes and the QHO is shown in Ref. \cite{Ole}, in terms of coordinates transformation. The important point is that the energy eigenstate of the QHO is isomorphic to the envelope field distribution of a HG mode. As a result, the propagation of the HG mode in free space changes this envelope focusing or diverging. Therefore, in order to control the evolution of QHO state by manipulating the HG mode, one uses the stroboscopic evolution. This means that we prepare a given state in the envelope distribution of the HG mode in a given position $z_{in}$, and propagate it through linear optical devices in order to obtain an evolved state at another position $z_{out}$. 

\begin{center}
\begin{figure}[h]
\includegraphics[width=2in,height=2.67in,keepaspectratio]{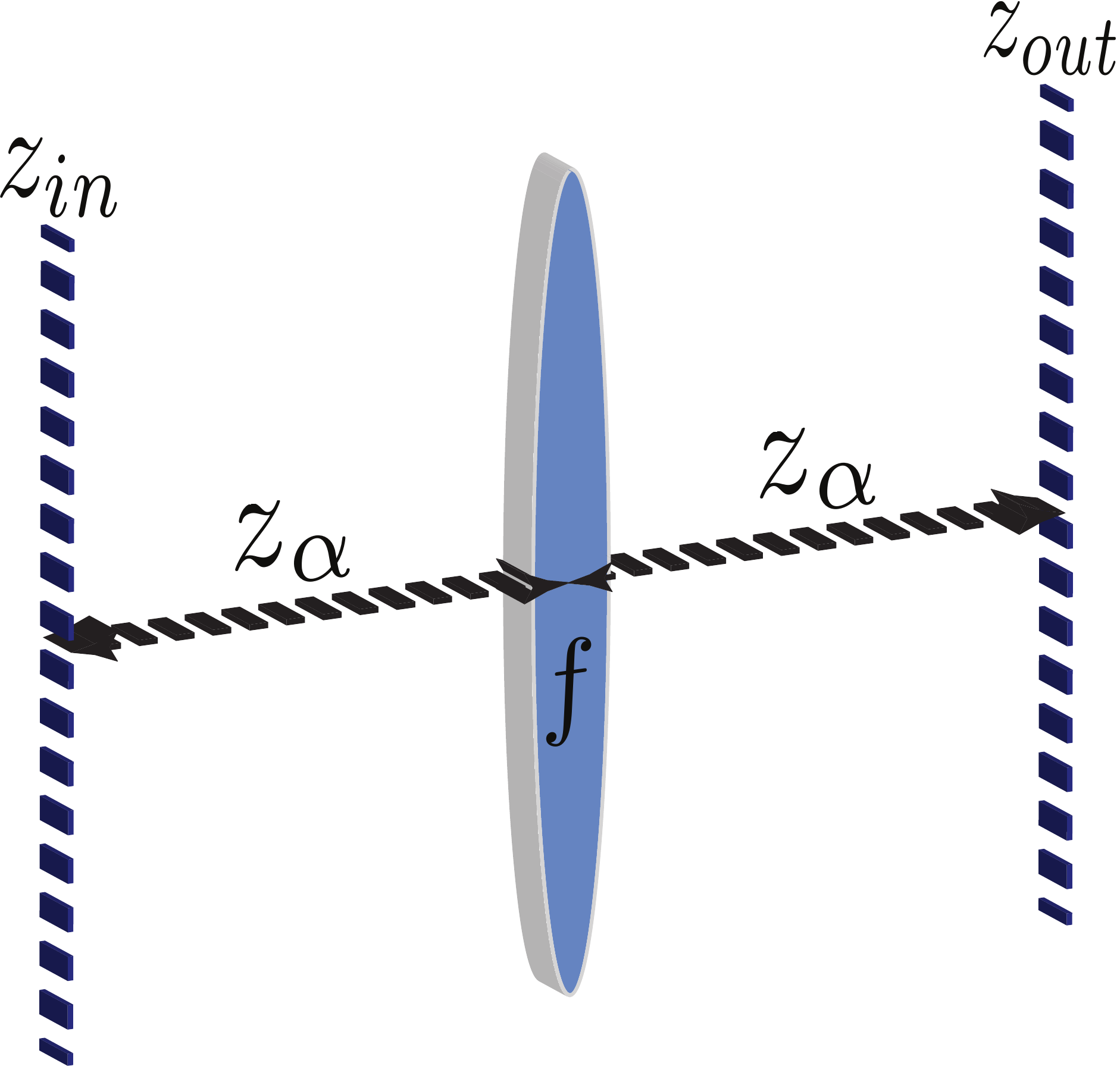}
\caption{(color online) Optical implementation of the fractional Fourier transform. A lens with focal length $f$ is placed equidistant between the input and output planes, located at $z_{in}$ and $z_{out}$, respectively. The distance $z_{\alpha}$ is related to the parameter $\alpha$ and to the focal length $f$ by $z_{\alpha} = 2f\sin^{2}(\alpha/2)$. The output field, $\Psi_{\alpha}(x,z_{out})$, is given by the fractional Fourier transform of the input field, $\Psi(x,z_{in})$, in the $x$ coordinate plane.}
\label{fig:fractional}
\end{figure}
\end{center}

The free evolution of the QHO is obtained by subjecting the HG mode to the so called Fractional Fourier Transform (FRFT). This is an integral transform that has found use in quantum mechanics \cite{wiener29,Namias,mcbride87,chountasis99}, signal processing, and optics \cite{pellat-finet94,Lohmann1,ozaktas01,tasca09a}, as it may be implemented in optical systems easily. The symmetric lens system shown in Fig. \ref{fig:fractional} realizes the optical FRFT. The field distribution at the input plane $\Psi(x, z_{in})$ is transformed into another distribution 
$\Psi_{\alpha}(x,z_{out})$ at
the output plane by means of free propagation, propagation through a lens, and another free propagation. It is characterized by a continuous order parameter $\alpha \in [0,2\pi]$ that is related to the distance
$z_{\alpha}$ by $z_{\alpha} = 2f\sin^{2}(\alpha/2)$, and can be defined as 
the operator \cite{pellat-finet94}
\begin{equation}
V_{\alpha} = e^{-i\alpha\frac{\mathbf{P}^{2} + \mathbf{X}^{2}}{2}},
\label{eq:frft}
\end{equation}
where $\mathbf{X}$ and $\mathbf{P}$ are the dimensionless position and momentum operators, respectively.

In other words, this operation corresponds to a rotation of the QHO by an angle $\alpha$ in the phase space, or a free evolution according to the QHO Hamiltonian. When compared to the optical Fourier Transform, the FRFT is a similar but a more general transformation, with the Fourier Transform being a special case with $\alpha = \pi/2$. For instance, in the experimental scheme of Fig. \ref{fig:fractional} one sets $z_{\alpha} = f$ to implement the optical Fourier Transform.

The mathematical statement of the action of the FRFT on the Hermite-Gaussian modes can be obtained by noticing that they represent the eigenmodes of the system Hamiltonian (the energy eigenstates), and are therefore the eigenfunctions of the $V_{\alpha}$ operator \cite{ozaktas01}
\begin{equation}
V_{\alpha} \phi_{n} = e^{-i \alpha \varepsilon_{n}} \phi_{n},
\label{eq:frft2}
\end{equation}
where $\phi_{n}(x) = \langle x\vert\phi_{n}\rangle$ is the position representation of the $n$-th eigenvector associated with the $n$-th eigenvalue $\varepsilon_{n}$ of the system Hamiltonian.

From the experimental point of view, the application of $V_{\alpha}$ on the eigen-mode $\phi_{n}$ encodes the information about the order of the mode $n$ (the system energy) in its optical phase. This is a crucial point for the measurement of the characteristic function. Moreover, the free parameter $\alpha$ can be experimentally controlled by varying $z_{\alpha}$ and $f$ (see Fig. \ref{fig:fractional}). 

%%%%%%%%%%%%%%%%%%%%%%%%%%%%%%%%%%%%%%%%%%%%%%%%%%%%%%%
%%%%%%%%%%%%%%%%%%%%%%%%%%%%%%%%%%%%%%%%%%%%%%%%%%%%%%%

\subsection{The optical interferometer}

\begin{figure}[h]
\centering
\includegraphics[width=3.5in,height=2.84in,keepaspectratio]{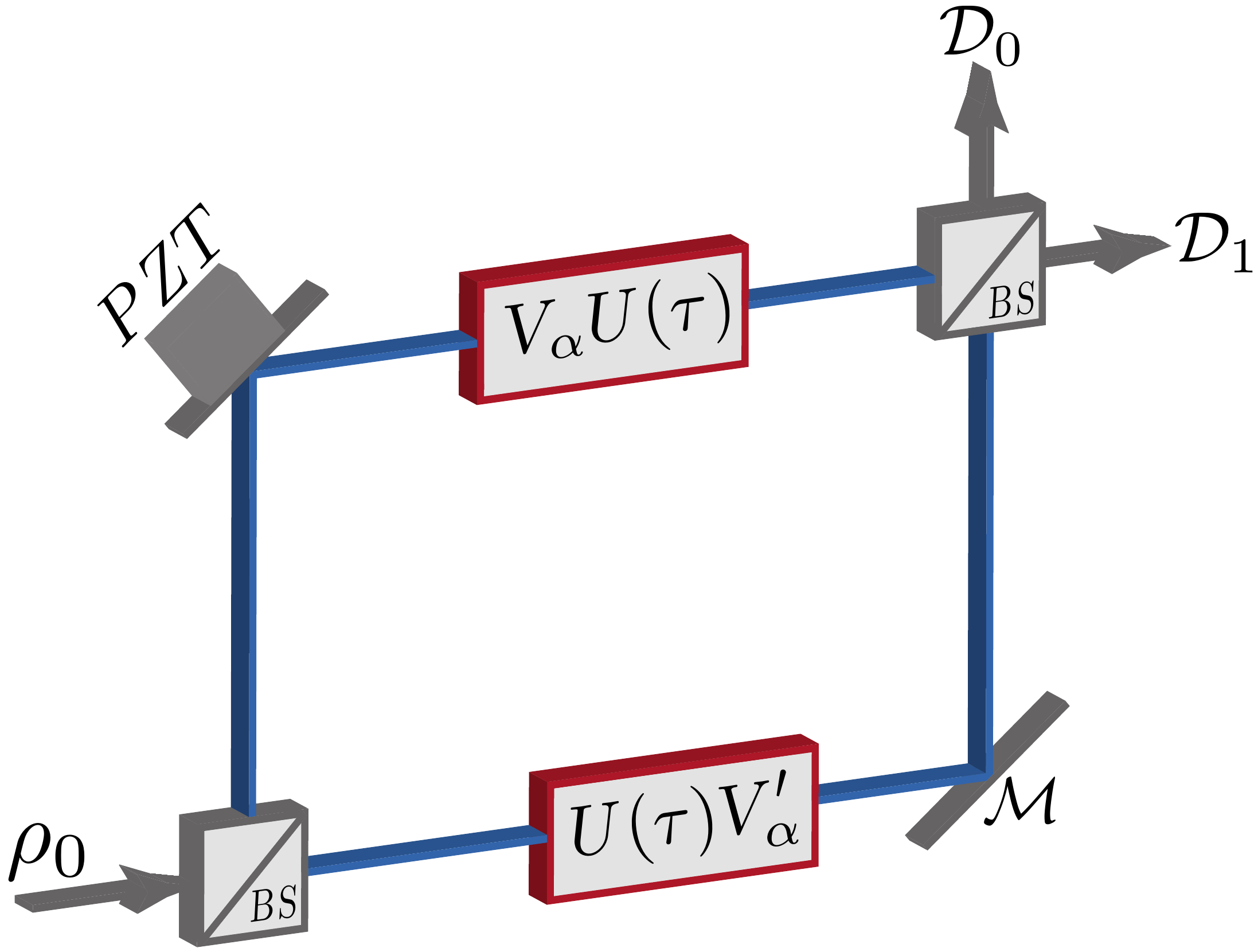}
\caption{(color online) Physical implementation of the protocol to measure the characteristic function. In the optical interferometer the input state is split. In the upper path it is imaged by a lens onto the FRFT device that realizes the free evolution $V_{\alpha}$ (see Fig. \ref{fig:fractional}). After the FRFT the process denoted by $U(\tau)$ is implemented and another lens images the final transverse distribution onto the output. In the lower path a similar scheme is realized, but it first goes through the process and then to the free evolution $V_{\alpha}'$ according to the new Hamiltonian. $\mathcal{M}$ is a mirror, $\mathcal{D}_{0}$ and $\mathcal{D}_{1}$ are bulk detectors, and PZT is used to control the phase difference and to alternate between the real and the imaginary parts of the characteristic function.}
\label{fig:set-up}
\end{figure}

Here we present a general discussion about the proposed set-up, leaving all the technical details to Appendix \ref{tech}.
Consider the interferometer sketched in Fig. \ref{fig:set-up}, which was built inspired on Refs. \cite{Mauro,Vlatko}. Let us first analyze it for one input mode $\phi_{n}^{I}$ prepared in a Hermite-Gaussian mode, which corresponds to an eigenstate of order $n$ of the system Hamiltonian. For simplicity, only one dimension $x$ is considered and therefore only one label for the modes is needed. However, the generalization for two dimensions is straightforward. 

The considered mode is split at the input of the interferometer, and in each path it undergoes distinct evolutions.

%%%%%%%%%%%%%%%%%%%%%%%%%%%%%%%%%%%%%%%%%%%%%%%%%%%%
%%%%%%%%%%%%%%%%%%%%%%%%%%%%%%%%%%%%%%%%%%%%%%%%%%%%%%%

In the upper path, we have the following transformation:
\begin{eqnarray}
\phi_{n}^{I} \rightarrow \mbox{Free evolution} \rightarrow  \phi_{n}^{I} \mbox{e}^{-i\varepsilon_{n}^{I}s}
\rightarrow \nonumber \\
\rightarrow \mbox{Process} \rightarrow  \mbox{e}^{-i\varepsilon_{n}^{I}s} \sum_m c_{m,n} \phi_{m}^{F}.
\label{eq:upper}
\end{eqnarray}
The phase factor after the second arrow appearing in Eq. (\ref{eq:upper}) follows from Eq. (\ref{eq:frft2}),
with $\alpha$ replaced with s. The coefficients $c_{m,n}$ describe the overlap between the input mode $n$ and the $m$-th component of the output mode, after the process, in terms of the final Hamiltonian basis. In order to implement the stroboscopic evolution, we use two additional lenses in each path. In the upper path, one of these lenses images the input plane of the interferometer onto the input plane of the FRFT device. The output plane of the FRFT is the input plane for the device realizing the process, and the other lens images the output plane of the process device onto the output plane of the interferometer. Similar procedure is followed in the lower path. The role of these lenses is to avoid free propagation of the optical modes, keeping the evolution of the spatial transverse distribution under control. The optical imaging realized by a lens reproduces the field distribution of the input plane in the output plane apart from a constant phase factor. This phase factor can be factored out. We are not accounting for it in the calculations because the total phase difference between the fields in the two paths can be controlled with the help of a piezoelectric actuator (PZT) in one of the mirrors. 

In the lower path the transformation is
\begin{eqnarray}
\phi_{n}^{I} \rightarrow \mbox{Process}  \rightarrow  \sum_m c_{m,n} \phi_{m}^{F}\rightarrow \nonumber \\
\mbox{Free evolution} \rightarrow \sum_m c_{m,n} \phi_{m}^{F}\mbox{e}^{-i\varepsilon_{m}^{F}s}.
\label{lower}
\end{eqnarray}
It is important to note here that the free evolution, in this last case, is generated by the final Hamiltonian, since this occurs after the process, which may change the system Hamiltonian. 

Taking into account the evolutions in both arms of the interferometer, the light intensity at the output is then proportional to
\begin{eqnarray}
I_{n} \propto 2A_{n} + Re \biggr\{ \sum_{m}  |c_{m,n}|^2 \, \mbox{e}^{i(\varepsilon_{m}^{F}-\varepsilon_{n}^{I})s}\biggr\}.
\end{eqnarray}
This result was obtained assuming that we have 50:50 beam splitters at the input and output and we have denoted by $A_{n}$, the intensity in each arm of the interferometer. 

We recall that the expansion coefficient
\begin{equation}\label{eq:coeff}
c_{m,n} = \int \int dx^{'} dx\,\, \left[ \phi_{m}^{F}(x^{'}) \right]^{\ast} U(x^{'},x,t) \phi_{n}^{I}(x),
\end{equation}
is the transition amplitude from mode $\phi_{n}^{I}(x)$ to mode $\phi_{m}^{F}(x)$ due to the action of the process,
written in terms of the functions describing the spatial transverse structure of the optical modes. 

As a final step, we sum up over the incoherent contributions of all modes composing the initial thermal state $\rho^{I}_{\mathcal{S}} = \sum_{n} p_{n}^{I} |\phi_{n}^{I}\rangle\langle\phi_{n}^{I}|$ resulting in
\begin{eqnarray}
I \propto 2A + Re \biggr\{  \sum_{m,n}p_{n}^{I}  |c_{m,n}|^2 \, \mbox{e}^{i(\varepsilon_{m}^{F}-\varepsilon_{n}^{I})s}\biggr\},
\label{c5}
\end{eqnarray}
where $A$ is the intensity in each path of the interferometer summing up over the contributions
of all input modes.

Comparing this result with the characteristic function in the form given by Eq. (\ref{eq:char11}), we can immediately see that the intensity $I$ is proportional to its real part $I \propto 2A + \mbox{Re}\left[G(s)\right]$, apart from the constant factor $2A$. In this calculation, we considered that the overall phase
difference between upper and lower paths was zero. However, the phase difference can be controlled using the PZT shown in Fig.
\ref{fig:set-up}, so that we can set it to $\pi/2$ in order to measure the imaginary part of characteristic
function. 

It is usual in optical interferometers to finely displace one of the mirrors with a PZT in order to observe oscillations of the output intensity and determine for instance, the visibility of the interference between the fields in the two paths. In the present application, the PZT is used to stabilize the phase difference in zero or $\pi/2$, and the oscillations of the output intensity are due to the variation of the parameter $\alpha$ of the free evolution $V_{\alpha}$ realized by the FRFT device, which is equivalent to varying the time of the free evolution.

The input state consists of a mixture of Hermite-Gaussian (HG) modes weighted by the Maxwell-Boltzmann distribution coefficients dependent on the temperature, or the coefficients $p_n$
of Eq. \ref{eq:prob-work}. However, as we have seen
in the analysis of the interferometer, we can make one experiment for each HG mode separately and register
the values of the output intensities for the real and imaginary parts. Having the data for all relevant 
modes, we can simply sum up all intensities with the corresponding Boltzmann weights for each mode. The Fourier analysis of this sum of intensities gives us the transition
probabilities and we can reconstruct the characteristic function. This procedure is valid because there is
no mutual coherence between two distinct input HG modes, as they belong to a thermal equilibrium state, which is a maximally mixed state. The relevant modes are those for which the coefficients $p_n$ are non-negligible at a given temperature. In practice, we define a cut-off value for the coefficient. Below the cut-off it is not necessary
to perform the experiment with the corresponding HG mode, because it practically does not contribute to the
work distribution.

\subsection{Example}

%%%%%%%%%%%%%%%%%%%%%%%%%%%%%%%%%%%%%%%%%%%%%%%%%%%%
\begin{figure}[h]
\centering
  % file name: C:/Users/PH/Desktop/COOPS/LUCAS/WORK/Versoes/ReplyPRA/prisma.pdf
  \includegraphics[width=3.59in,height=1in,keepaspectratio]{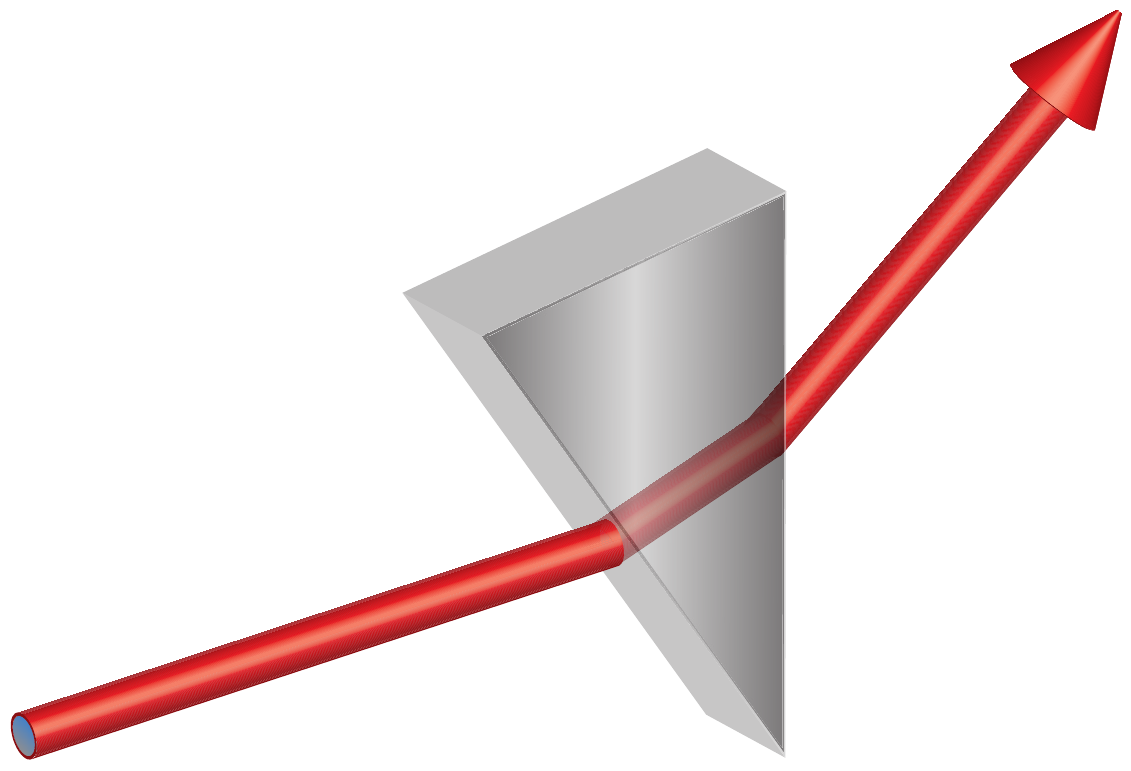}
   \caption{(color online) Refraction through a prism. The change in the direction of propagation
realizes the displacement operator for the transverse momentum.}
\label{fig:prisma}
\end{figure}
%%%%%%%%%%%%%%%%%%%%%%%%%%%%%%%%%%%%%%%%%%%%%%%%%%%%%%%

In previous sections, we showed that the optical interferometer can be used to measure the characteristic function for an arbitrary process acting on a QHO and then to reconstruct the associated work distribution. Now, as an illustrative example we choose a particular process which displaces the linear momentum of the oscillator by $p_{0}$. This can be achieved experimentally by a proper refraction of the light beam. See Fig. \ref{fig:prisma} for a possible simple implementation. In this case, the initial and final Hamiltonians in the Schr\"{o}dinger picture, are given by
\begin{equation}
\mathbf{H}_{I} = \frac{\mathbf{P}^{2}}{2m} + \frac{m\omega^{2}}{2}\mathbf{X}^{2},
\label{hamil_i}
\end{equation}
and
\begin{equation}
\mathbf{H}_{F} = \frac{\left( \mathbf{P} + p_{0} \right)^{2}}{2m} + \frac{m\omega^{2}}{2}\mathbf{X}^{2},
\label{hamil_f}
\end{equation}
where $\mathbf{P}$ and $\mathbf{X}$ are the linear momentum and position operators respectively. As the initial and final Hamiltonians are connected by a similarity transformation $\mathbf{H}_{F} = D^{\dagger}(p_{0}) \mathbf{H}_{I}D(p_{0})$, they have the same energy spectrum
\begin{equation*}
\varepsilon^{F}_{n} =\varepsilon^{I}_{n} \equiv \varepsilon_{n} = \hbar\omega\left(n + \frac{1}{2}\right), \hspace{1cm} n = 0, 1, 2, ...,
\end{equation*}
with the displacement operator $D(p_{0})$ being defined as 
\begin{equation}
D(p_{0}) \equiv e^{-ip_{0}\mathbf{X}/\hbar}.
\label{eq:process}
\end{equation}
The eigenvectors of $\mathbf{H}_{I}$ are Fock states $| \phi_{n}^{I} \rangle = | n \rangle$ (Hermite-Gaussian modes in the space representation), while the eigenvectors of $\mathbf{H}_{F}$ are $| \phi_{n}^{F} \rangle = D^{\dagger}(p_{0})| n \rangle $. The evolution between such Hamiltonians is made by a sudden quench 
\cite{messiah}, which means that $\lim_{\delta t \rightarrow 0}U(t+\delta t,t) = \textbf{1}$. Therefore, the coefficient (\ref{eq:coeff}) reduces to 
\begin{equation}\label{eq:c_mn}
c_{m,n}=\int dx\phi_{m}^{I}(x)\phi_{n}^{I}(x)e^{-ip_{0}x/\hbar}, 
\end{equation}
whose expression is developed in the Appendix \ref{harmonic}. Considering this process we can compute the characteristic function as

%%%%%%%%%%%%%%%%%%%%%%%%%%%%%%%%%%%%%%%%%%%%%%%%%%%%%%%%%%
\begin{figure*}[ht]
\centering
\includegraphics[width=3.3in,height=1.69in,keepaspectratio]{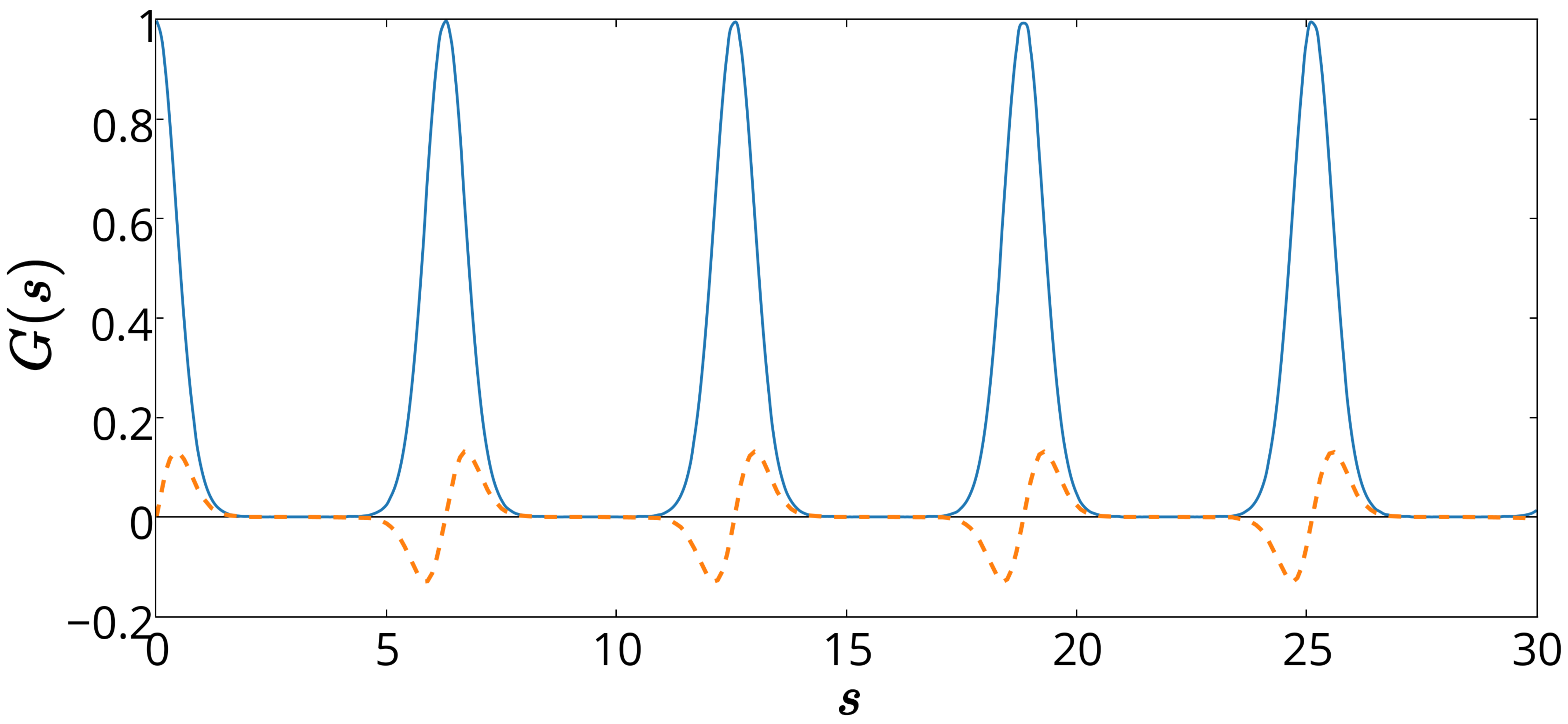} \hspace{0.2cm} \includegraphics[width=3.3in,height=1.69in,keepaspectratio]{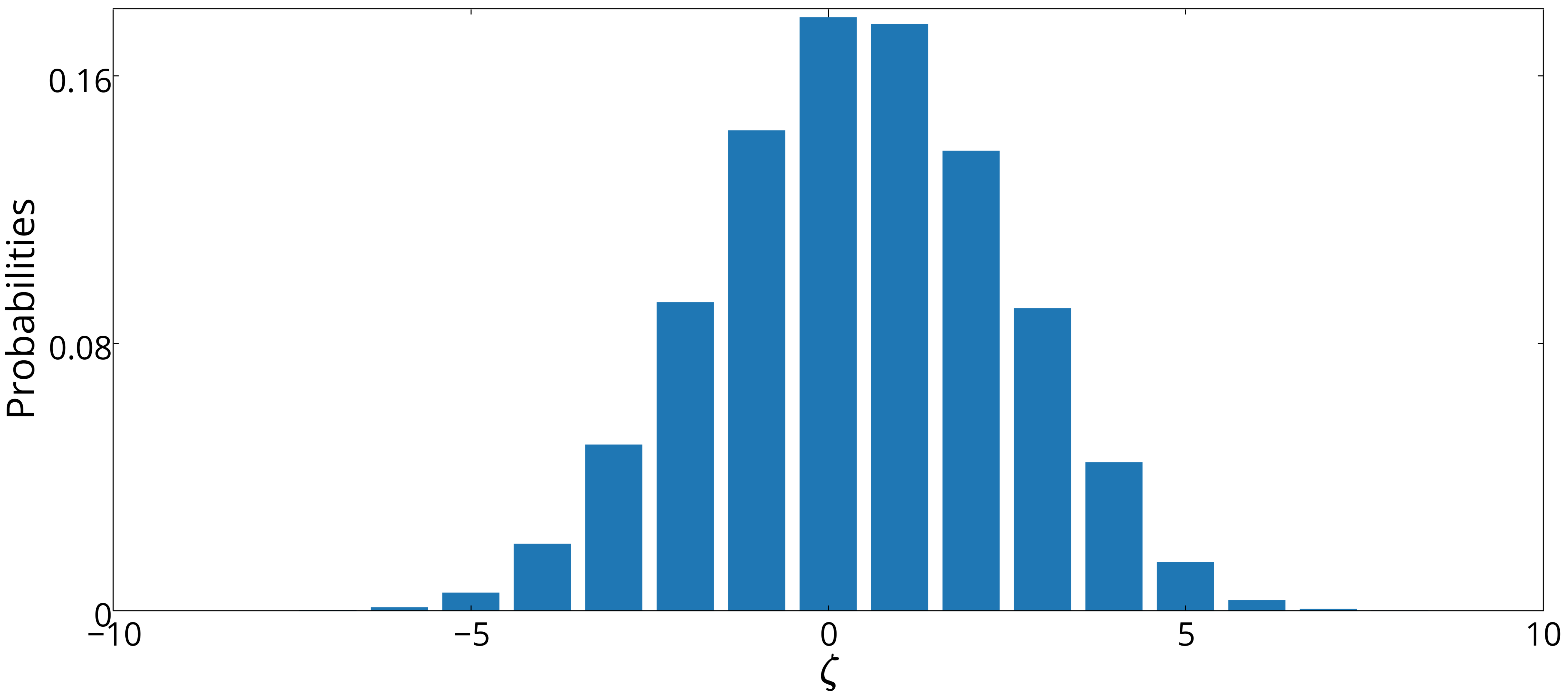}\\
\includegraphics[width=3.3in,height=1.69in,keepaspectratio]{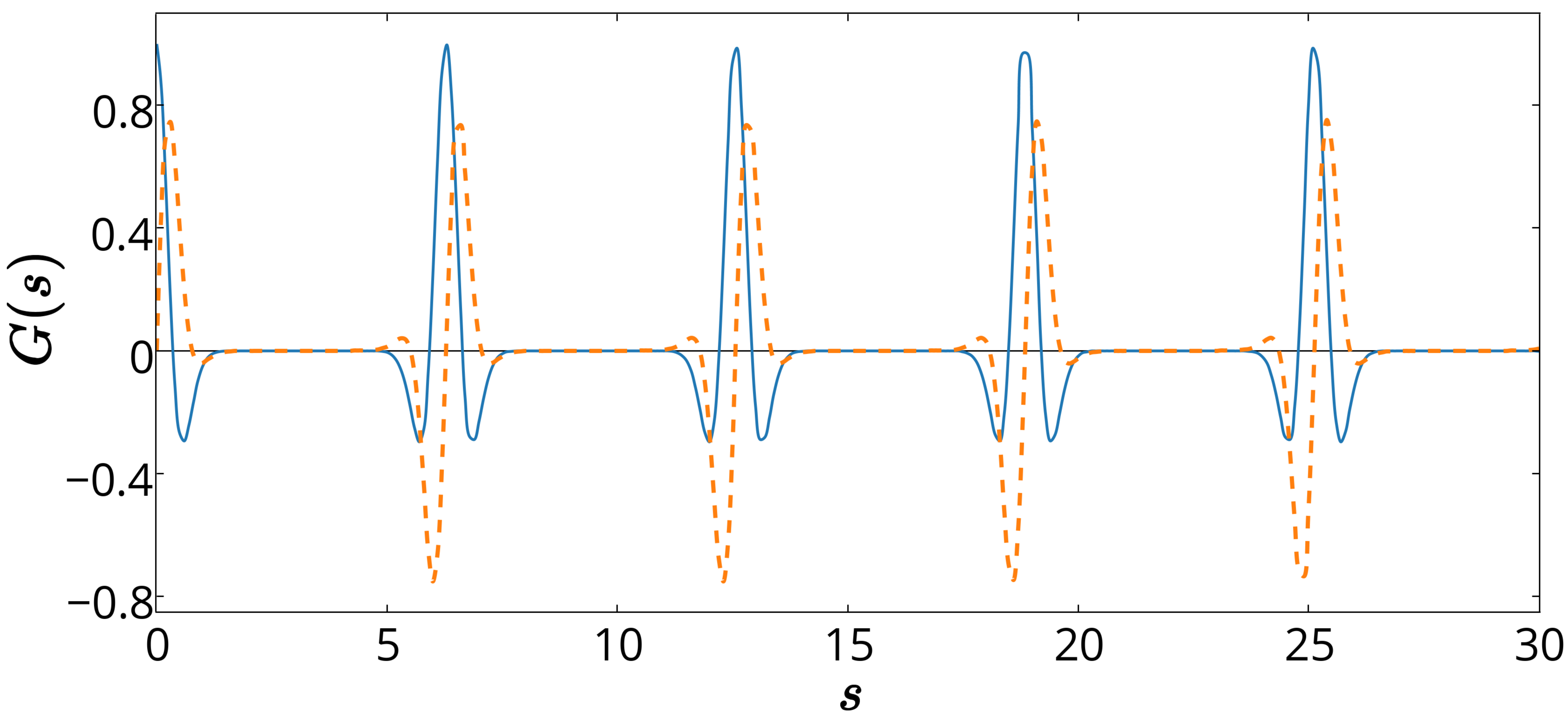} \hspace{0.2cm} \includegraphics[width=3.3in,height=1.69in,keepaspectratio]{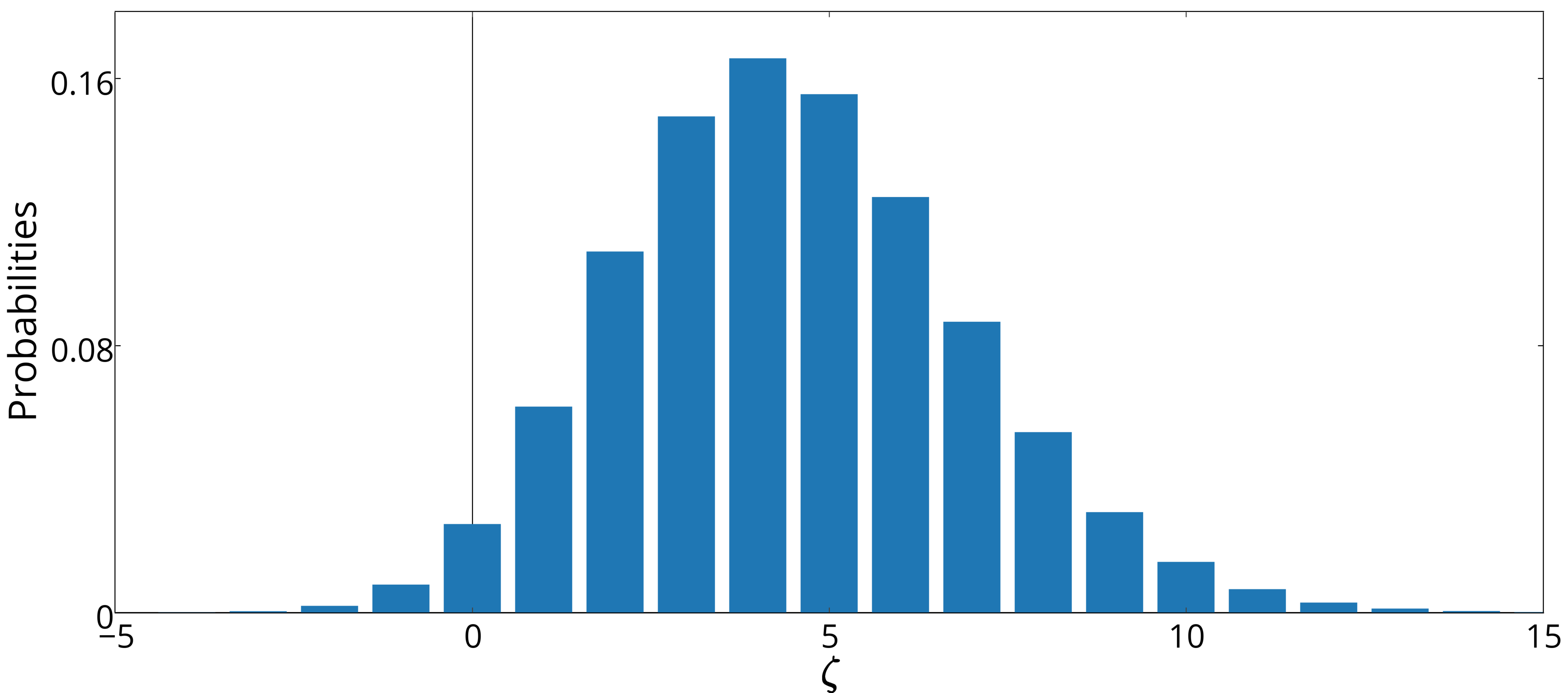}
\caption{(color online) Characteristic function (left column) and the probabilities associated with the work distribution (right column) for the kicked harmonic oscillator. The parameters are $q_0 = 1$ and $\hbar\omega/kT = 0.1$ in the top figures and $q_0 = 3$ and $\hbar\omega/kT = 1$ for the bottom figures. The solid and dashed lines in the left column are the real and imaginary parts of the characteristic function, respectively.}
\label{fig:plots}
\end{figure*}
%%%%%%%%%%%%%%%%%%%%%%%%%%%%%%%%%%%%%%%%%%%%%%%%%%%%%%%%%%%%

\begin{eqnarray}
G\left(s\right) &=& \sum_{m,n}\frac{p_{n}^{I}q_{0}^{2(m+n)}e^{-q_{0}^{2}/2}}{2^{m+n}n!m!}e^{is(m - n)} \nonumber\\
&\times &\left\vert\sum_{r=0}^{\min(m,n)}r!2^{r}\binom{m}{r}\binom{n}{r}(-iq_{0})^{-2r}\right\vert^{2},
\label{eq:char_f}
\end{eqnarray}
from which follows the work distribution
\begin{eqnarray}
P\left(\zeta\right) &=&  \sum_{m,n}\frac{p_{n}^{I}q_{0}^{2(m+n)}e^{-q_{0}^{2}/2}}{2^{m+n}n!m!}\delta\left[\zeta - (m - n)\right] \nonumber\\
&\times &\left\vert\sum_{r=0}^{\min(m,n)}r!2^{r}(-iq_{0})^{-2r}\binom{m}{r}\binom{n}{r}\right\vert^{2},
\label{eq:work_f}
\end{eqnarray} 
where $P\left(\zeta\right)$ is obtained by simply Fourier transforming $G\left(s\right)$, and $\zeta = \mathcal{W}/\hbar\omega$ is a dimensionless quantity measuring work in units of $\hbar\omega$, while $q_{0}=p_{0}/\sqrt{m\omega\hbar}$ is a dimensionless scale for the quench. $m$ and $\omega$ characterize the oscillator.

Fig. \ref{fig:plots} shows the behavior of the characteristic function and the work distribution for some values of $q_0$ and for some temperatures of the initial thermal state expressed in terms of the ratio $\hbar\omega/kT$.

\section{Experimental viability}

The proposed experiment is rather simple and allows a high degree of control over the parameters of the system and process. This is because the use of interferometers and the optical tools necessary for the implementation are very well developed. A detailed description of one possible implementation is given in Appendix \ref{tech}.  However, there some technical challenges that must be overcome in a real set. The first one and probably the hardest is the implementation and control of the FRFT inside the interferometer. The control parameter $\alpha$ must be varied in order to obtain the desired oscillations whose frequencies encode the energy difference of the transitions induced by the process. This can be accomplished by varying the focal length $f$ of the lens and the propagation distances $z_{\alpha}$ (See Fig. \ref{fig:fractional} for the optical 
implementation of the FRFT) before and after it simultaneously, taking care to respect the relation $z_{\alpha} = 2f\sin^{2}(\alpha/2)$ between them. Even though lenses with variable focal lengths are commercially available, the simultaneous control of the focal length and free propagation distances resulting in a controlled variation of the FRFT order $\alpha$ has not yet been reported as far as we know.   

Another difficulty concerns the stabilization of the phase difference of the interferometer in zero or $\pi/2$ and the proper production and alignment of Hermite-Gaussian modes through the interferometer. This type of optical operation has already been accomplished in several set-ups, even though not combined with the FRFT control.

Therefore, despite some technical issues, the experiment is clearly realizable with current technology.

\section{Discussion}
   
We now discuss some physical aspects of the set-up, demonstrating its generality and usefulness for the study of thermodynamics of quantum systems.

{\em Versatility of the set-up.} Although we concentrate in the harmonic oscillator case in the text, our scheme is not limited to this Hamiltonian. It relies on the isomorphism between the the non-relativistic quantum dynamics of a particle under the action of a given potential and the paraxial Helmholtz equation for the light propagating in a medium with modulated index of refraction (see Eqs. (\ref{eq:helmholtz})). Therefore, by suitably changing the index of refraction $\Delta n(x)/n_{0}$ we can modify the effective potential of the analog quantum system. An immediate and simple variation is the two-dimension
quantum harmonic oscillator. For a proper phase difference between the oscillations in two orthogonal axis,
we get eigenstates described by the Laguerre-Gaussian optical modes. Interestingly, they are also eigenfunctions of the FRFT shown in Fig. \ref{fig:fractional}, so that exactly the same set-up could be used
to measure the characteristic function associated with some work performed on this system.

{\em Open dynamics.} Since every system interacts with its environment, the study of general processes, described by completely positive and trace preserving (CPTP) maps, becomes very important to boost both theoretical and technological developments in this field. From the thermodynamics point of view, when considering open systems, it is not always possible to distinguish between work and heat. However, fluctuation relations still hold and we can address questions like how entropy is produced in the system of interest or what are the role played by correlations in thermodynamic processes. Moreover, as the technological developments are driving us through the path of miniaturization, the interesting question of how non-Markovian evolutions modify the irreversible properties of the dynamical system is an important subject to be experimentally addressed.

We show in Appendix \ref{App_open} that the proposed optical scheme can be used to study open dynamics. The only modification required is the inclusion of an ancillary degree of freedom playing the role
of the environment. Using the polarization, for instance, we can even choose between tracing over the
environmental degrees of freedom or measuring them. As dissipative systems can be modeled by certain non-Hermitian Hamiltonians \cite{Bender}, they could also be investigated with this experimental system.

{\em Quantum versus classical.} In this proposal, we are concerned with the quantum harmonic oscillator. Its dynamics is emulated by the stroboscopic evolution of light modes that are analog to the QHO energy eigenfunctions. However, the whole experiment is realized with classical light, for instance a laser beam
prepared in the HG modes. Going to the photon counting regime would not change the results of the experiment, because the degree of freedom related to the QHO dynamics is the spatial mode and not the energy (number of photons) of the light field. We recall that the final measurement is an intensity measurement
at the output of the interferometer, and it is subjected to noise due to photon number fluctuations. However,
this is not a fundamental aspect, as this noise can be made to tend to zero as the intensity tends to
infinity.

From the point of view of the emulation of the QHO, we could go from the quantum to the classical regime by increasing the temperature of the initial state. In this case the ratio $\hbar\omega/kT \rightarrow 0$, meaning that the separation between the energy levels would be very small compared to the thermal energy, and the work distribution would tend to a continuous distribution. In Ref. \cite{PRX_Jarzynski} this limit is analyzed for the forced quartic harmonic oscillator. Interesting features of the quantum-classical transition
were already apparent dealing with $n$ = 150 energy states. We consider that it is very important to perform experiments to test the boundaries between the quantum and the classical worlds,
because there is not a general method to perform it theoretically. The experiment to study the
quantum to classical transition of a chaotic system also realized with the spatial degrees of freedom of
light is an example of this type of investigation \cite{Gabi}.

Inspired by these ideas, let us assume that emulating energy states with $n \simeq$ 150 for the QHO in our proposal would also be enough to study interesting aspects of the quantum-classical transition. The technical requirement for this emulation is the ability of generating high order Hermite-Gaussian modes, like a HG$_{150,150}$ for instance. A device for generating this mode must have enough spatial resolution. Recalling that a HG$_{n,n}$ mode has $(n+1)^{2}$ light spots in its transverse profile, we can estimate 151$^{2}$ = 22801 light spots in the transverse distribution of a HG$_{150,150}$ mode. If we use for instance a 4K resolution spatial light modulator to generate this mode, there is a matrix containing 4094$\times$2464 = 10087616 pixels (10Mp) available, which correspond to $\simeq$ 442 pixels/light spot. A more concrete image for this scheme is to think that we can count on 22801 squares of $\sqrt{\mbox{442}} \simeq 21$ pixels size for modulating some input wavefront to generate a HG$_{150,150}$ mode. Therefore, the realization of such high order modes is clearly feasible with commercial devices. From this rough estimation, we can conclude that the interferometric set-up we propose here can be very helpful in this kind of investigation. 

{\em Definition of work.} Considering unitary processes, there are several distinct definitions of work in the literature \cite{Allahverdyan,Roncigli,Deffner,Pigeon,Horodecki,Deffner1}. Due to its interferometric character, our setup was designed to study the two-time measurement definition of work. Modifications of the proposed setup would eventually allow the experimental study of these different definitions, as well as the changes introduced in the fluctuation relations and in the quantum-to-classical transition. However, this possibility must be analyzed case by case and we leave this study for future works.

\section{Conclusion}

In summary, we demonstrated that an optical interferometer can be used to measure the characteristic function associated with a given process implemented in an optical system for both unitary and non-unitary dynamics. We show that it emulates a quantum harmonic oscillator and it may be
modified in order to emulate other interesting systems. Several processes can be easily implemented with linear optical devices. This set-up is feasible with current technology and represents a valuable platform for advancing the understanding and testing experimentally the quantum limits of thermodynamics.

\section{Acknowledgments}

This work was funded by the Brazilian funding agencies CNPq (Grants No. 401230/2014-7, 445516/2014-3 and 305086/2013-8), CAPES, and the National Institute for Quantum Information (INCT-IQ).

\bibliographystyle{apsrev}

\appendix

%%%%%%%%%%%%%%%%%%%%%%%%%%%%%%%%%%%%%%%%%%%%%%%
\section{Technical details of the proposed set-up}
\label{tech}
\begin{widetext}
%%%%%%%%%%%%%%%%%%%%%%%%%%%%%%%%%%%%%%%%%%%%%%%%%%%%
\begin{figure*}[h]
\centering

  % file name: C:/Users/PH/Desktop/COOPS/LUCAS/WORK/Versoes/ReplyPRA/setup.pdf
  \includegraphics[bb=0 47 842 595,width=6in,height=5in,keepaspectratio]{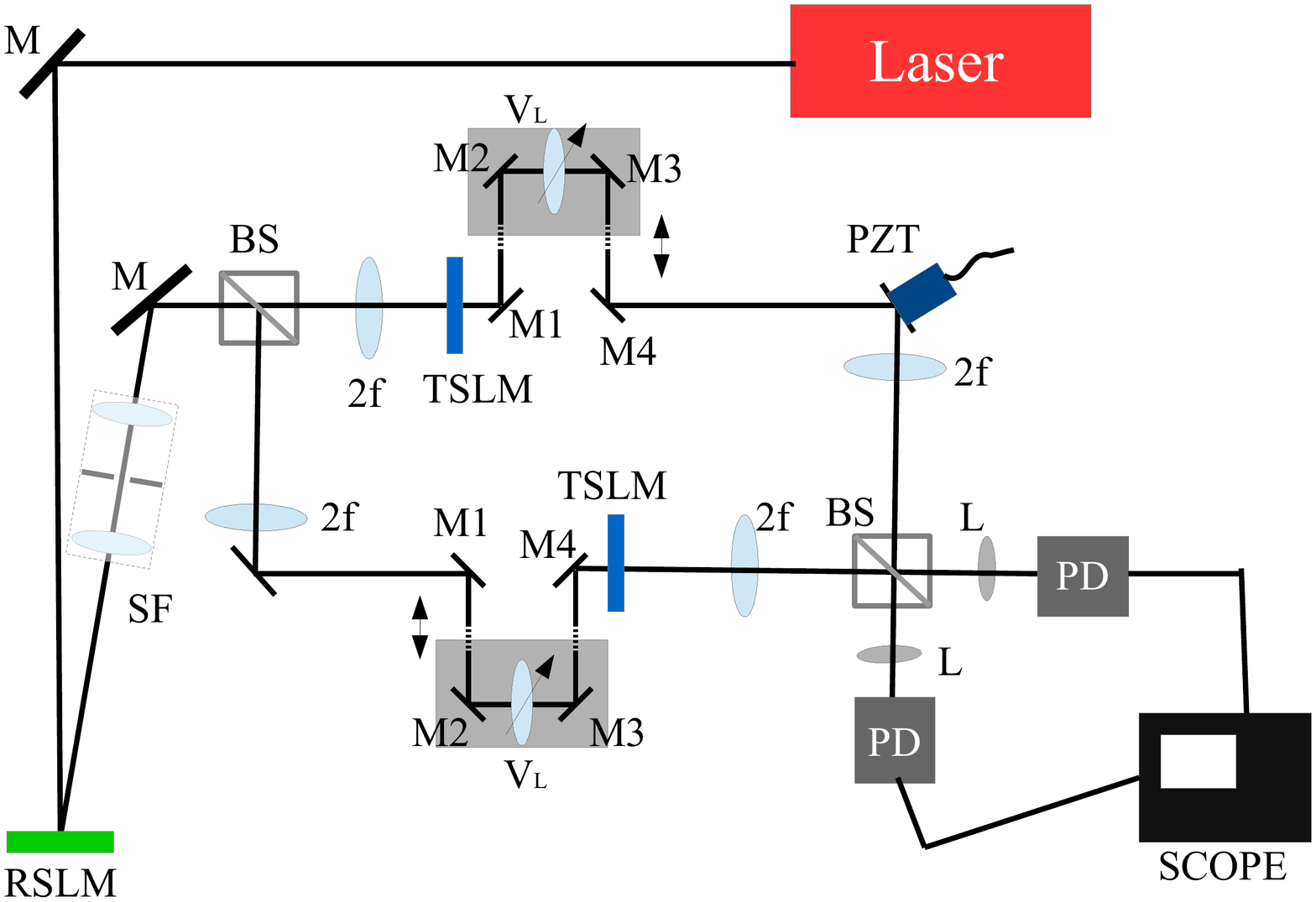}

\caption{(color online) Experimental set-up. M is mirror, RSLM is reflection spatial light
modulator, SF is spatial filter, BS is beam splitter, 2f is a lens used in an imaging configuration,
TSLM is transmission spatial light modulator, V$_L$ is variable lens, PZT is piezoelctric actuator,
L is lens, PD is photodiode, and SCOPE is a digital oscilloscope. See more details in the text. }
\label{fig:detail}
\end{figure*}
%%%%%%%%%%%%%%%%%%%%%%%%%%%%%%%%%%%%%%%%%%%%%%%%%%%%%%%
\end{widetext}
Let us consider the experimental set-up in a more concrete basis, explaining the
working principle of all building blocks of Fig. \ref{fig:set-up}. A complete version of
the set-up is shown in Fig. \ref{fig:detail}. It is usually convenient
to use a laser as the light source, specially in an interferometric set-up. A Helium-Neon
laser is a good option, as it is generally easy to find He-Ne lasers with a good spatial
mode, which allows a better control in the production of Hermite-Gaussian (HG) modes. The laser
beam is then sent to a reflective spatial light modulator (RSLM), where the HG modes are prepared.
Manufacturers as Holoeye, for instance, offer 4k resolution panel SLMs. Therefore, the spatial
resolution can be very high, allowing the production of high order HG modes.

After the RSLM, the HG modes pass through a spatial filter in order to eliminate eventual
noise in the transverse spatial profile, usually due to the pixelation of the RSLM.
The HG modes are sent through the interferometer by splitting in a 50/50 beam splitter. 
In the upper path there is a $2f$ lens system projecting the input profile onto the
transmission spatial light modulator (TSLM). The $2f$ lens system is just a lens placed
at a symmetric position in between object and image plane, at a distance equal to two times
the focal length from them. This results in the imaging of the object with magnification 1.
The fact that the image is inverted is compensated by the existence of two $2f$ systems
in each path of the interferometer. 

The TSLM can implement a large variety of unitary processes, like the displacement
analyzed in the illustrating example, by simply applying proper transverse phase distributions. 
After the TSLM, the light beam immediately enters the system that implements a controllable Fractional 
Fourier Transform (FRFT).
This system consists of four mirrors and a variable lens. 
M$_1$ is the input mirror, and M$_2$ is the output mirror. 
Inside, there are two other mirrors, M$_3$ and M$_4$, and a variable lens V$_l$ mounted on a translation 
stage, so that the path from mirror M$_1$ to V$_l$ and from V$_l$ to M$_4$ can be carefully adjusted.
The proper and combined control of these distances and the focal length of V$_l$ allows the realization of
the FRFT with variable and controlled order. Lenses with variable focal length are commercial products
and are manufactured for instance by Optotune.

After going through the FRFT evolution, the light beam is reflected by a mirror mounted
on a PZT actuator. We can use it to finely tune the phase difference between the two optical
paths of the interferometer. This is useful for instance, to set the proper phase difference between
the two outputs of the interferometer in order to actually have the real and imaginary parts
of the characteristic function. Following this mirror, there is another $2f$ lens system, which
images the output of the FRFT system to the output beam splitter of the interferometer.

In the lower path, the optical devices are similar. The only difference is that the FRFT
is realized before the process. The two output beams of the interferometer are focused on
photodiodes and both currents are registered by a digital oscilloscope. It accumulates data
from the runs with all relevant HG modes and these data can be treated to find the characteristic
function spectrum. 

\section{The work distribution for the harmonic oscillator under displacement process}
\label{harmonic}

As stated in the main text, to calculate the work distribution first we obtain the characteristic function (\ref{eq:char11}). The quantities appearing in $G(s)$ are the eigenvalues of the initial and final Hamiltonians, respectively,
\begin{equation}
\mathbf{H}_{I} = \frac{\mathbf{P}^{2}}{2m} + \frac{m\omega^{2}}{2}\mathbf{X}^{2},
\label{app:hamilA}
\end{equation}
and
\begin{equation}
\mathbf{H}_{F} = \frac{\left( \mathbf{P} + p_{0} \right)^{2}}{2m} + \frac{m\omega^{2}}{2}\mathbf{X}^{2},
\label{hamil_fA}
\end{equation}
which are equals and given by
\begin{equation*}
\varepsilon_{n}^{0} = \varepsilon_{n}^{\tau} \equiv \varepsilon_{n}=\hbar\omega\left(n + \frac{1}{2}\right), \hspace{1cm} n = 0, 1, 2, ...,
\end{equation*}
where $\mathbf{P}$ and $\mathbf{X}$ are the momentum and position operators, and  $p_{0}$ is the displacement of the linear momentum of the oscillator. The other quantity necessary to evaluate the characteristic function is $p_{m,n}=e^{-\beta\varepsilon_{n}} \vert c_{m,n} \vert^{2} /Z_{I}$, where $\beta$ is the inverse temperature, $Z_{I}$ is the partition function, and the coefficients $c_{m,n}$ are described in Eq. (\ref{eq:c_mn}), i.e.,
\begin{equation}\label{eq:c_mnA}
c_{m,n}=\int dx\phi_{m}^{I}(x)\phi_{n}^{I}(x)e^{-ip_{0}x/\hbar}.
\end{equation}
The eigenfunctions of the Hamiltonian $\mathbf{H}_{I}$ are given by
\begin{equation*}
\phi_{n}^{I}(x) \equiv \langle x\vert n \rangle = \left(\frac{\gamma}{\pi}\right)^{\frac{1}{4}}\frac{1}{\sqrt{2^{n}n!}}e^{-\frac{\gamma x^{2}}{2}}\mathcal{H}_{n}\left(\sqrt{\gamma}x\right),
\end{equation*}
where $\gamma \equiv m\omega/\hbar$ and $\mathcal{H}_{n}$ are the Hermite polynomials generated by the recursion relation
\begin{equation*}
\mathcal{H}_{n}(x) = (-1)^{n}e^{x^{2}}\frac{d^{n}}{dx^{n}}\left(e^{-x^{2}}\right).
\end{equation*}
This leads us to
\begin{equation*}
c_{m,n} = a_{mn}\int dx e^{-ip_{0}x/\hbar}e^{-\gamma x^{2}}\mathcal{H}_{m}\left(\sqrt{\gamma}x\right)\mathcal{H}_{n}\left(\sqrt{\gamma}x\right),
\end{equation*}
with $a_{mn} = \left(\gamma/\pi2^{m+n} n!m!\right)^{1/2}$. It is convenient to work with dimensionless variables and to do this we define $\zeta = \sqrt{\gamma}x$, obtaining
\begin{equation*}
c_{m,n} = \frac{a_{mn}}{\sqrt{\gamma}}\int d\zeta e^{-iq_{0}\zeta}e^{-\zeta^{2}}\mathcal{H}_{m}\left(\zeta\right)\mathcal{H}_{n}\left(\zeta\right),
\end{equation*}
with $q_{0} \equiv p_{0}/\hbar\sqrt{\gamma}$ a dimensionless quantity.

Now, using the relation
\begin{equation*}
\mathcal{H}_{m}\left(\zeta\right)\mathcal{H}_{n}\left(\zeta\right) = \sum_{r=0}^{\min(m,n)}r!2^{r}\binom{m}{r}\binom{n}{r}\mathcal{H}_{m+n-2r}(\zeta),
\end{equation*}
with 
\begin{equation*}
\binom{m}{r} = \frac{m!}{r!(m-r)!} \hspace{0.5cm} \mbox{for} \hspace{0.5cm} 0\leq r \leq m,
\end{equation*}
we obtain
\begin{eqnarray*}
c_{m,n} &=& \frac{a_{mn}}{\sqrt{\gamma}}\sum_{r=0}^{\min(m,n)}r!2^{r}\binom{m}{r}\binom{n}{r}\nonumber \\
&\times & \int d\zeta e^{-iq_{0}\zeta}e^{-\zeta^{2}}\mathcal{H}_{m+n-2r}(\zeta).
\end{eqnarray*}
Let us concentrate on the integral that appeared in this last equation. By noting that $l\equiv m+n-2r$ is always positive or zero and also that
\begin{equation*}
\mathcal{H}_{l}(\zeta) = 2^{l/2}\mathcal{H}_{el}\left(\sqrt{2}\zeta\right),
\end{equation*}
where $\mathcal{H}_{el}$ is the modified Hermite polinomial of order $l$, the above integral can be rewritten as
\begin{equation*}
\int d\zeta e^{-iq_{0}\zeta}e^{-\zeta^{2}}\mathcal{H}_{l}(\zeta) = \frac{2^{l/2}}{\sqrt{2}}\int d\zeta e^{-i\frac{q_{0}}{\sqrt{2}}\zeta}e^{-\zeta^{2}/2}\mathcal{H}_{el}(\zeta).
\end{equation*}

Now, we multiply the generating function of $\mathcal{H}_{el}$
\begin{equation*}
e^{\zeta t - t^{2}/2} = \sum_{l}\mathcal{H}_{el}(\zeta)\frac{t^{l}}{l!}
\end{equation*}
by $e^{-\zeta^{2}/2}$ and take the Fourier transform (from variable $\zeta$ to the variable $q_{0}$) of the result, thus obtaining
\begin{eqnarray*}
\mathcal{F}\left[e^{\zeta t - t^{2}/2 - \zeta^{2}/2}\right] &=& e^{-q_{0}^{2}/2 - itq_{0}} \nonumber\\
&=&e^{-q_{0}^{2}/2}\sum_{n}\frac{t^{l}}{l!}(-iq_{0})^{n}\nonumber\\
&=& \sum_{l}\mathcal{F}\left[e^{-\zeta^{2}/2}\mathcal{H}_{el}(\zeta)\right]\frac{t^{l}}{l!}.\nonumber
\end{eqnarray*}
By equating the same powers on $t$ we obtain
\begin{equation*}
\mathcal{F}\left[e^{-\zeta^{2}/2}H_{el}(\zeta)\right] = e^{-q_{0}^{2}/2}(-iq_{0})^{l},
\end{equation*}
which leads us to
\begin{eqnarray*}
c_{m,n} &=& \frac{(-iq_{0})^{m+n}e^{-q_{0}^{2}/4}}{\sqrt{2^{m+n}n!m!}}\sum_{r=0}^{\min(m,n)}r!2^{r}\binom{m}{r}\binom{n}{r}(-iq_{0})^{-2r}.
\end{eqnarray*}

Therefore, the characteristic function and work distribution (the inverse Fourier transform of the characteristic function) can be written as
\begin{widetext}
\begin{equation}
G\left(s\right) = \sum_{m,n}\frac{p_{n}^{I}q_{0}^{2(m+n)}e^{-q_{0}^{2}/2}}{2^{m+n}n!m!}\left\vert\sum_{r=0}^{\min(m,n)}r!2^{r}\binom{m}{r}\binom{n}{r}(-iq_{0})^{-2r}\right\vert^{2}e^{is(m - n)}.
\label{char_f}
\end{equation}

\begin{equation}
P\left(\zeta\right) =  \sum_{m,n}\frac{p_{n}^{I}q_{0}^{2(m+n)}e^{-q_{0}^{2}/2}}{2^{m+n}n!m!}\left\vert\sum_{r=0}^{\min(m,n)}r!2^{r}(-iq_{0})^{-2r}\binom{m}{r}\binom{n}{r}\right\vert^{2}\delta\left[\zeta - (m - n)\right],
\label{work_f}
\end{equation}
\end{widetext} 
where $\zeta = \mathcal{W}/\hbar\omega$ is a dimensionless quantity measuring work in units of $\hbar\omega$.

\section{Open processes}
\label{App_open}

\begin{figure}[h]
\centering
\includegraphics[width=3.5in,height=2.84in,keepaspectratio]{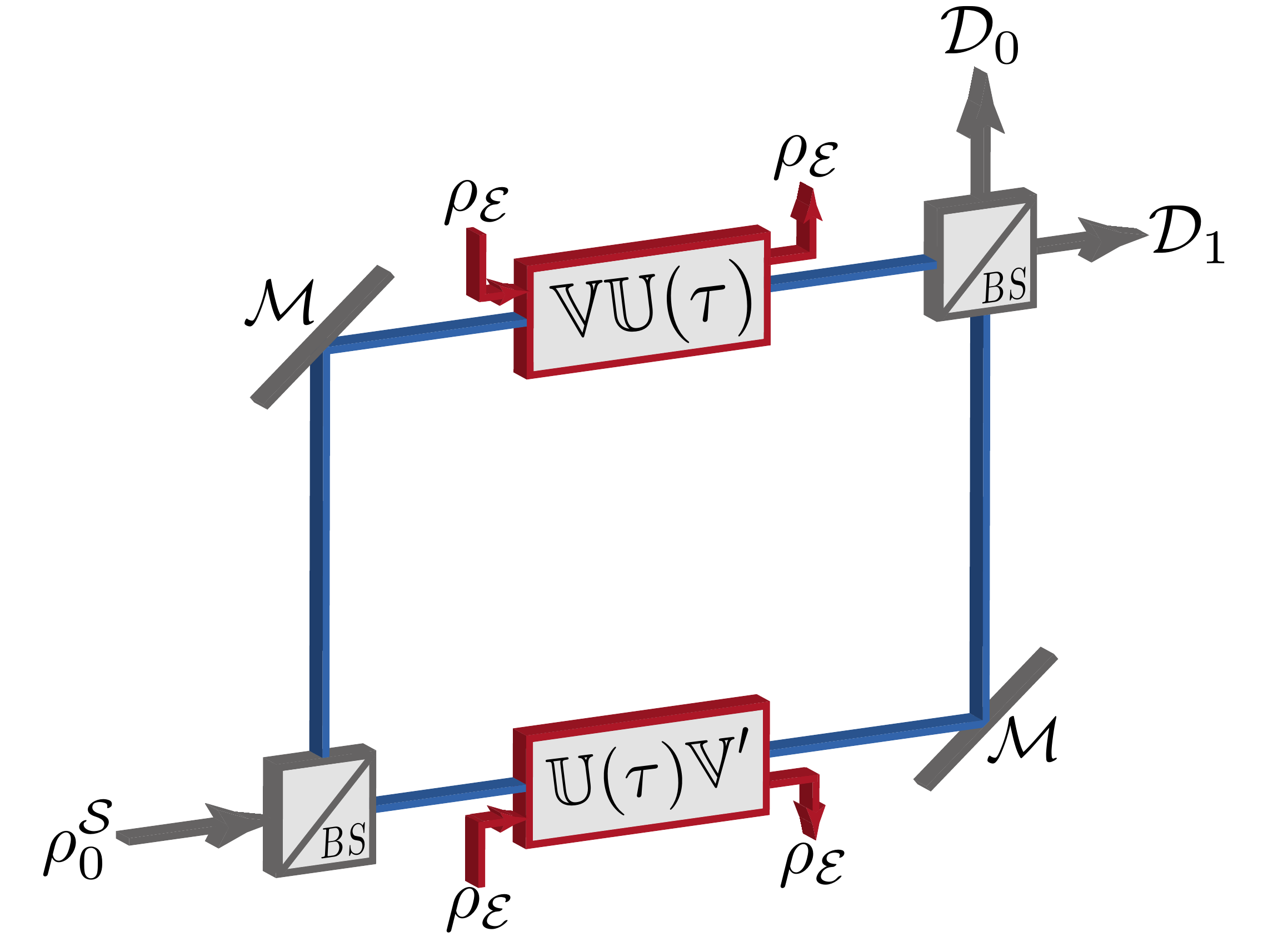}
\caption{(color online) Optical interferometric implementation of the protocol to measure the characteristic function for an open dynamics. The paths of the interferometer are the auxiliary system ($\mathcal{A}$), labeled as $|1\rangle_{\mathcal{A}}$ (upper) and $|0\rangle_{\mathcal{A}}$ (lower). $\mathbb{U}(\tau)$ is the process while $\mathbb{V}$ and $\mathbb{V'}$ are related to the free evolution of the system (see text for more details). Both of these operators act on the joint Hilbert space of the system ($\mathcal{S}$) and environment ($\mathcal{E}$), whose state is $\rho_{\mathcal{E}}$.}
\label{fig:int_open}
\end{figure}

We now generalize our set-up in order to study open processes, including Markovian and non-Markovian dynamics.
See Fig. \ref{fig:int_open}.
 
The most general evolution of a quantum system, initially in state $\rho_{0}$, is mathematically described by a completely positive and trace preserving (CPTP) map $\Phi$, which can be written in the operator-sum representation as
\begin{equation}
\rho_0 \rightarrow \Phi(\rho_0) = \sum_{m}\Gamma_{m}\rho_0\Gamma_{m}^{\dagger},
\label{eq:open_evol}
\end{equation}
with $\sum_{m}\Gamma_{m}^{\dagger}\Gamma_{m} = \mathds{1}$. $\Gamma_{m}$ are the so called Krauss operators and will be defined bellow. If $\sum_{m}\Gamma_{m}\Gamma_{m}^{\dagger} = \mathds{1}$ holds, then the identity is preserved and the map is called unital.
 
In Ref. \cite{Deffner}, the authors derived a general fluctuation relation valid for any CPTP map and for any pair of initial and final operators. We show now that the interferometric scheme we described can be used in order to address this case too. 

An energy measurement is performed on the system, resulting in the outcome $u_{n}^{I}$ ($H_{I} = \sum_{m}u^{I}_{m}\Pi_{m}^{I}$ is the initial Hamiltonian). After this, a general process $\Phi$ is applied to the system and a second energy measurement, with outcome $u^{F}_{m}$ is performed. The probability distribution for the random variable $u_{mn} = u_{m}^{F} - u_{n}^{I}$ is then
\begin{equation}
P(u) = \langle\delta(u - u_{mn})\rangle,
\end{equation}
with the joint probability given by 
\begin{equation}
p_{mn} = \mbox{Tr}\left\lbrace\Pi_{m}^{F}\Phi\left[\Pi_{n}^{I}\rho_0\Pi_{n}^{I}\right]\right\rbrace.
\end{equation}
$\Pi_{n}^{I}$ and $\Pi_{n}^{F}$ are the eigenmatrices of the initial and final Hamiltonians, respectively.

From this we can define the associated characteristic function
\begin{eqnarray}
G(s) &=& \int du P(u)e^{isu} \nonumber \\
&=& \mbox{Tr}\left\lbrace V_{F}^{\dagger}\Phi\left[M_{I}(\rho_0)V_{I}\right]\right\rbrace,
\label{eq:open_char}
\end{eqnarray}
with $V_{I(F)}$ being the initial (final) free evolutions of the system and
\begin{equation}
M_{I}(\rho_0) = \sum_m \Pi_{m}^{I}\rho_0\Pi_{m}^{I}.
\end{equation}

By choosing $s = i\beta$ we obtain the fluctuation relation \cite{Deffner}
\begin{equation}
\langle e^{-\beta\,u} \rangle = \gamma,
\label{eq:open_jar}
\end{equation} 
with
\begin{equation}
\gamma = \mbox{Tr}\left\lbrace(V_F)^{\dagger}\Phi\left[V_I\, M_{I}(\rho_0)\right]\right\rbrace .
\label{eq:open_gamma}
\end{equation}

It is important to observe here that $u$ in Eq. (\ref{eq:open_jar}) cannot be directly identified as work, since the system is open and we also have heat. For unitary evolutions we obtain the usual fluctuation relation.

\begin{figure}[h]
\centering
\includegraphics[width=3in,height=2in,keepaspectratio]{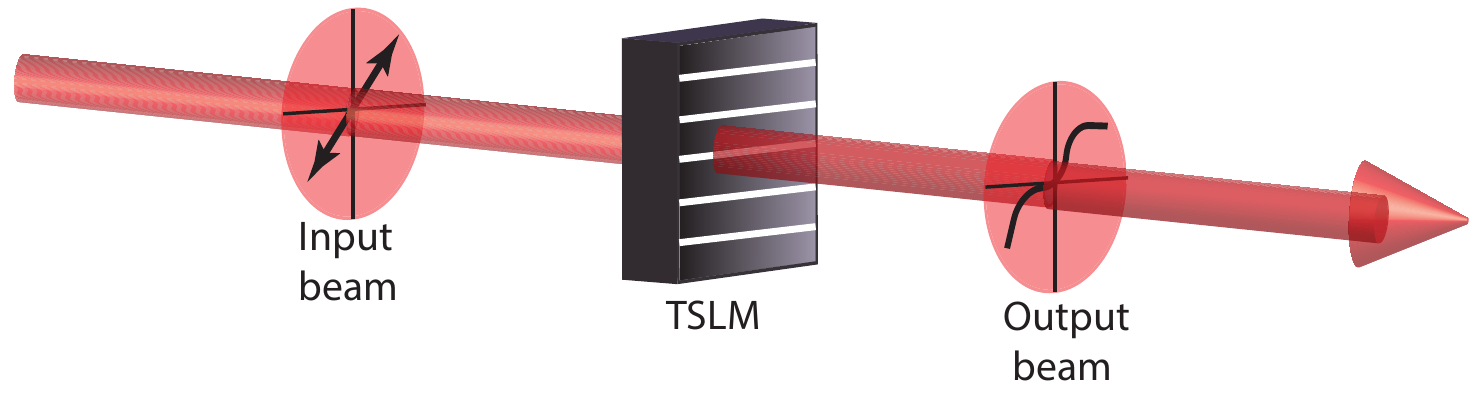}
\caption{(color online) Example of implementation of a non-unitary process.}
\label{open}
\end{figure}

The only modification that should be introduced in the set-up in order to experimentally study this general fluctuation relation concerns the way one implements the process. One example of possible non-unitary process is sketched in Fig. \ref{open}. The light beam, for instance one HG mode, is prepared in a diagonal
linear polarization state. The process is implemented by a transmission spatial light modulator (TSLM),
which modulates the phase of the horizontal polarization component, but not the vertical one. This is
a typical characteristic of liquid crystal SLMs. After being modulated only in the horizontal polarization
component the output beam proceeds in the usual way. This is valid for both arms of the interferometer. 
If one traces out the polarization degree of freedom, classically speaking, if one performs the detection
without any kind of polarization selection, then there are two effects. One is the reduction of the 
electric field amplitude participating to the process in comparison to what one would have if the whole
beam was modulated. Another effect is the noise induced by the polarization component that was not
modulated. It will contribute to the intensity signal measured, but it carries no information about
the system and the process. 

The use of the polarization as the environmental degree of freedom has the advantage that
one could eventually perform polarization dependent measurements in order to try to retrieve
the information about the effect of the environment. These are ideas that must be further
developed. Our point is just to demonstrate the potential of the experimental scheme as
platform to study the quantum limits of thermodynamics.

%%%%%%%%%%%%%%%%%%%%%%%%%%%%%%%%%%%%%%%%%%%%%%%%%%%%
%%%%%%%%%%%%%%%%%%%%%%%%%%%%%%%%%%%%%%%%%%%%%%%%%%%%%%%

First, let us assume that the state of the environment is a pure state, represented by $|\xi_{\mathcal{E}}\rangle$. Since purification does not change the physics of the system, this does not impose any restriction on the set-up. Following the same procedure explained in the last section, we start with the system in the pure state $|\phi_{\mathcal{S}}^{n}\rangle$, which is an eigenstate of the initial Hamiltonian. Therefore, the complete initial state (ancilla $\mathcal{A}$ plus system $\mathcal{S}$ and environment $\mathcal{E}$) is given by $|\psi_{\mathcal{ASE}}^{I}\rangle = |0_{\mathcal{A}},\phi_{\mathcal{S}}^{n},\xi_{\mathcal{E}}\rangle$. The final state, at the output of the interferometer, will be given by
\begin{eqnarray}
\rho_{\mathcal{ASE}}^{F} &=& \frac{1}{4}\left[|0_{\mathcal{A}}\rangle\langle 0_{\mathcal{A}}|\otimes|\chi_{\mathcal{SE}}^{+}\rangle\langle \chi_{\mathcal{SE}}^{+}| \right. \nonumber\\
&+& \left. |1_{\mathcal{A}}\rangle\langle 1_{\mathcal{A}}|\otimes|\chi_{\mathcal{SE}}^{-}\rangle\langle \chi_{\mathcal{SE}}^{-}| \right. \nonumber\\
&+& \left. |0_{\mathcal{A}}\rangle\langle 1_{\mathcal{A}}|\otimes|\chi_{\mathcal{SE}}^{+}\rangle\langle \chi_{\mathcal{SE}}^{-}| \right. \nonumber\\
&+&\left. |1_{\mathcal{A}}\rangle\langle 0_{\mathcal{A}}|\otimes|\chi_{\mathcal{SE}}^{-}\rangle\langle \chi_{\mathcal{SE}}^{+}| \right],
\end{eqnarray}
where
\begin{equation}
|\chi_{\mathcal{SE}}^{\pm}\rangle = \left[\mathbb{V'U} \pm \mathbb{UV}\right]|\phi_{\mathcal{S}}^{n},\xi_{\mathcal{E}}\rangle. 
\end{equation}

Now, by taking the trace over the system and environment, we get the reduced density matrix of the ancilla.
\begin{eqnarray}
\rho_{\mathcal{A}} &=& \mbox{Tr}_{\mathcal{SE}}\rho_{\mathcal{ASE}}^{F} \\
&=& \frac{1}{2}\left[ \mathds{1}_{\mathcal{A}} + \mbox{Re}\left\lbrace\mbox{Tr}\hat{O}_{\mathcal{SE}}\right\rbrace\sigma_{z} + \mbox{Im}\left\lbrace\mbox{Tr}\hat{O}_{\mathcal{SE}}\right\rbrace\sigma_{y}\right],\nonumber
\label{eq:reduced_open}
\end{eqnarray}
with $\sigma_i$ being the $i$-th Pauli matrix while
\begin{equation}
\hat{O}_{\mathcal{SE}} = \left(\mathbb{V'U}\right)^{\dagger}\mathbb{UV}\rho_{\mathcal{SE}},
\end{equation}
and $\rho_{\mathcal{SE}} = |\phi_{\mathcal{S}}^{n},\xi_{\mathcal{E}}\rangle\langle\phi_{\mathcal{S}}^{n},\xi_{\mathcal{E}}|$. 

We proceed by computing the trace appearing in Eq. (\ref{eq:reduced_open}). In order to do this, we assume that $\mathbb{V} = V_{\mathcal{S}}\otimes\mathds{1}_{\mathcal{E}}$, with a similar definition for $\mathbb{V'}$. In this definition, $V_{\mathcal{S}}$ is the free evolution of the system of interest (the same operator appearing in the closed case explained in the last section). This means that the desired process and the interaction with the environment will be taken into account in the definition of $\mathbb{U}$. Since $\mathbb{U}$ is completely general, this assumption does not impose any additional restriction to the set-up. Taking these considerations into account, we can write 
\begin{equation}
\mbox{Tr}\hat{O}_{\mathcal{SE}} = \mbox{Tr}_{\mathcal{S}}\left[(V')^{\dagger}\mbox{Tr}_{\mathcal{E}}\left[\mathbb{U}|\xi_{\mathcal{E}}\rangle\left(V\rho_{\mathcal{S}}^{0}\right)\langle\xi_{\mathcal{E}}|\mathbb{U}^{\dagger}\right]\right].
\end{equation}
Choosing a specific basis $\left\lbrace|\zeta^{m}_{\mathcal{E}}\rangle\right\rbrace$ for the environment we can compute the trace as
\begin{equation}
\mbox{Tr}\hat{O}_{\mathcal{SE}} = \mbox{Tr}_{\mathcal{S}}\left[(V')^{\dagger}\sum_{m}\langle\zeta_{\mathcal{E}}^{m}|\mathbb{U}|\xi_{\mathcal{E}}\rangle\left(V\rho_{\mathcal{S}}^{0}\right)\langle\xi_{\mathcal{E}}|\mathbb{U}^{\dagger}|\zeta_{\mathcal{E}}^{m}\rangle\right].
\end{equation}
But $\Gamma_{m} = \langle\zeta_{\mathcal{E}}^{m}|\mathbb{U}|\xi_{\mathcal{E}}\rangle$ is precisely the definition of the $m$-th Krauss operator. This explicitly indicates the non-uniqueness of the decomposition appearing in Eq. \ref{eq:open_evol}, since it depends on the choice of the basis for the environment. However, the physical evolution, i.e. the final state of the system, does not change. From this we obtain
\begin{equation}
\mbox{Tr}\hat{O}_{\mathcal{SE}} = \mbox{Tr}_{\mathcal{S}}\left[(V')^{\dagger}\Phi\left(V\rho_{\mathcal{S}}^{0}\right)\right],
\end{equation}
which is exactly the characteristic function given in Eq. (\ref{eq:open_char}) considering the case of an eigenvector of the initial Hamiltonian as the initial state of the system. This result has the same structure of the unitary case \cite{Mauro,Vlatko}, but now we have the action of the map on the system. Therefore, we can employ the same interferometric scheme of the last section in order to study non-unitary processes acting on the system.

\end{document}